\documentclass[journal]{IEEEtran}
\usepackage{amsmath,amssymb,amsfonts,mathtools}
\usepackage{algorithm}
\usepackage{algpseudocode}
\usepackage{graphicx}
\usepackage{array}
\usepackage{stfloats}
\usepackage{cite}
\usepackage{url}
\usepackage{textcomp}
\usepackage{booktabs}
\usepackage{capt-of}
\usepackage{siunitx}
\usepackage{tikz}
\usepackage{xcolor}
\usetikzlibrary{positioning,arrows.meta}
\usepackage[caption=false,font=footnotesize]{subfig}
\usepackage{placeins}

\definecolor{filippo}{rgb}{1,0.8,0.1}

\definecolor{mario}{rgb}{0.8,0.8,1}

\begin{document}
\bstctlcite{IEEEexample:BSTcontrol}
\title{Real-Time Coordination of Autonomous Vehicles at Intersections}
\author{Saugat Shahi, Jonathan Frey, Filippo Fabiani, Moritz Diehl and Mario Zanon}

\maketitle

\begin{abstract}
When coordinating connected and automated vehicles at an unsignalized intersection, the crossing order and the vehicle trajectories must be decided jointly. For a fixed crossing order, the problem can be formulated conveniently in a bi-level form where each vehicle solves its own optimal control problem, whose optimal value is a function of its entry and exit times, and a coordinator optimizes these times subject to precedence constraints. We compute the first and second order sensitivities of lower-level problems, which are required to solve the upper-level problem, by means of algorithmic differentiation using the real-time software package \texttt{acados}. This implementation exploits tailored algorithms to obtain a computationally efficient implementation of the optimal coordination algorithm. We validate our implementation in numerical simulations that demonstrate both real-time feasibility and effectiveness of our algorithm.
\end{abstract}

\section{Introduction}
\label{sec:introduction}

At an unsignalized intersection, connected and automated vehicles (CAVs) travel through the conflict zone directly, which can reduce delay, energy consumption and collision risk \cite{riostorres_survey,campos_mag}. Coordinating them requires two decisions: a discrete crossing order and the continuous trajectories optimization. Together they form a mixed-integer nonlinear program (MINLP) \cite{hult_theory,fayazi}, and deciding optimal crossing order is combinatorially hard \cite{colombo}. Existing work therefore fix the schedule, such as first-come-first-served reservation \cite{dresner_stone} or static priorities \cite{qian_priority}, but their solutions far from optimal for heterogeneous fleets \cite{shahi2026lagrange}.

For a fixed crossing order, the problem can be decomposed into an upper-level nonlinear program (NLP) that assigns occupancy slots of the intersection and independent optimal control problems (OCPs), one per vehicle \cite{hult_theory,hult_primal,hult2020}. The semi-distributed sequential quadratic programming (SQP) and interior-point schemes of \cite{hult2020,hult_ip} build on this structure and solve only a small quadratic program (QP) centrally. Similar formulations for a fixed schedule are found in \cite{malikopoulos_auto,zhang_cassandras,malikopoulos_time,kim_kumar,campos_rhc,katriniok,xu_distributed,murgovski,qian_mpc,kneissl,chalaki}.

Choosing the crossing order remains the primary challenge. Exact methods formulate it with integer variables and use mixed-integer programming \cite{fayazi,ge_graphical,hult_miqp}, tree search \cite{xu_tree}, branch-and-bound \cite{li_zhou} or dedicated sequencing algorithms \cite{yan}, but scales poorly. Heuristics instead improve a given order by swapping adjacent vehicles \cite{hult_miqp,faris_reorder}. In our previous study, the Lagrange multipliers associated with the precedence constraints quantify the cost related to the crossing order. They rank the critical vehicle pairs discussed in \cite{shahi2026lagrange}, which retrieves a significant portion of the optimality gap found in first-in-first-out. The extension in \cite{shahi2026mpc} incorporates this prioritization into a model predictive control (MPC) framework \cite{rawlings_mpc} that retains a history of performed swaps. Nevertheless, each swap is verified through a complete re-solving with a general-purpose NLP solver \cite{ipopt,casadi}, which prevents realtime operation as the number of vehicles increases.

Parametric sensitivity analysis is an alternative to handle these re-solves. Under standard assumptions, the solution of a parametric NLP is differentiable in the parameter, and its derivatives follow from the Karush--Kuhn--Tucker (KKT) system \cite{fiacco,buskens_maurer}. A second-order solver factorizes this system anyway, so first- and second-order derivatives of the optimal value come at little extra cost \cite{buskens_maurer,zavala_biegler}. The same idea underlies real-time iteration schemes \cite{rti,rawlings_mpc} and the solution sensitivities of \texttt{acados} \cite{acados,frey2025differentiable}, which builds on \texttt{HPIPM} \cite{hpipm} and \texttt{CasADi} \cite{casadi}.

This paper cast the coordination problem of \cite{shahi2026lagrange,shahi2026mpc} in bi-level form, in which each vehicle solves its own \texttt{acados} OCP and its optimal value is a function of its crossing times. Our contributions are fourfold. First, we compute the gradient of these value functions from the multipliers of the crossing conditions and their Hessian from the \texttt{acados} solution sensitivities. Second, we use them in a prediction QP that estimates the cost change of a candidate swap. Only swaps predicted to improve are confirmed by a re-solve, which replaces most of the full re-solves of \cite{shahi2026lagrange,shahi2026mpc}. Third, we solve the upper level by SQP with a dense or a structured \texttt{HPIPM} solver. Fourth, we compare the solutions of centralized and decentralized formulation and evaluate real time feasibility in closed loop simulation with computation delay against controllers in section~\ref{sec:res_closed}.

Section~\ref{sec:formulation} presents the centralized formulation and the multiplier-guided schedule refinement, Section~\ref{sec:decentralized} the decentralized formulation and the sensitivity-based method, Section~\ref{sec:results} the numerical results, and Section~\ref{sec:conclusion} concludes the paper.

\subsubsection*{Notation}
We denote $\mathbb{I}_a^b \coloneqq \{a,\ldots,b\}$, with $\mathbb{I}_a^b=\emptyset$ if $a>b$, and by $|\mathcal{S}|$ the cardinality of a finite set $\mathcal{S}$. The sampling time is $t_\mathrm{s}$, $k$ is the sampling instant, $N$ the prediction horizon and $n\in\mathbb{I}_0^N$ the time step, which refers to the time $(k+n)t_\mathrm{s}$. The placeholders $\dagger,\ddagger\in\{\mathrm{in},\mathrm{out}\}$ denote entry and exit quantities. For vehicle $i$, $x_i\coloneqq(x_{i,0},\ldots,x_{i,N})$, $u_i\coloneqq(u_{i,0},\ldots,u_{i,N-1})$ and $t_i\coloneqq(t_i^{\mathrm{in}},t_i^{\mathrm{out}})$. A \emph{schedule} $S$ is a permutation of the vehicles, $S_c$ is the vehicle in position $c$, and $\mathrm{swap}(S,c)$ exchanges positions $c$ and $c+1$. The superscripts $(j)$ and $\star$ denote the $j$-th SQP iterate and optimal quantities, and the subscript $\mathrm{sw}$ a swapped schedule. We use $\partial$ for partial and $\mathrm d$ for total derivatives, and $\operatorname{diag}(A_i)_{i\in\mathcal{I}}$ the block-diagonal matrix with blocks $A_i$.

\begin{figure}[t]
\centering
\includegraphics[width=\linewidth]{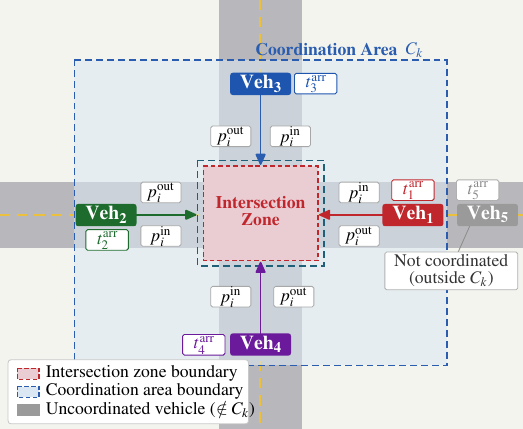}
\caption{Four-way unsignalized intersection scenario. Vehicles $\mathrm{Veh}_1$--$\mathrm{Veh}_4$ arrive asynchronously from different directions at times $t_1^{\mathrm{arr}},\ldots,t_4^{\mathrm{arr}}$ and travel along fixed straight paths toward the intersection zone (dashed red boundary). Entry and exit positions are denoted by $p_i^\mathrm{in}$ and $p_i^\mathrm{out}$, respectively, and colored arrows indicate each travelling direction. Vehicle $\mathrm{Veh}_5$ has not yet entered the coordination area $\mathcal{C}_k$ (dashed blue boundary) and is therefore not coordinated.}
\label{fig:intersection_diagram}
\end{figure}
\section{Problem formulation}
\label{sec:formulation}
We consider the coordination of cooperative CAVs approaching an unsignalized intersection, as illustrated in Fig.~\ref{fig:intersection_diagram}, following the setting of \cite{shahi2026lagrange,shahi2026mpc}. Each vehicle $i$ moves along a fixed path, with state $x_{i,n}$ and control input $u_{i,n}$ at time step $n$. The dynamics are obtained from a twice continuously differentiable continuous-time model by discretization with sampling time $t_\mathrm{s}$ and a zero-order hold on the input over each sampling interval:
\begin{equation}
    x_{i,n+1}=f_i(x_{i,n}, u_{i,n}),
    \label{eq:dyn}
\end{equation}
subject to path constraints
\begin{equation}
    h_i(x_{i,n},u_{i,n})\le 0.
    \label{eq:path}
\end{equation}
Vehicle $i$ enters and exits the intersection at fixed positions $p_i^{\mathrm{in}}$ and $p_i^{\mathrm{out}}$, with entry and exit times implicitly defined via
\begin{align*}
    p_i\left(t_i^{\dagger},x_i,u_i\right)=p_i^{\dagger},
\end{align*}
where $p_i(t,x_i,u_i)$ is the position of vehicle $i$ at time $t$. Since $x_i$ and $u_i$ are sequences defined at the sampling instants only, $p_i$ is evaluated between them by holding the input $u_{i,n}$ constant over the sampling interval $[(k+n)t_\mathrm{s},(k+n+1)t_\mathrm{s})$ and propagating the continuous-time model from $x_{i,n}$ over the fraction of the interval elapsed at time $t$; in this way, the pair $(x_i,u_i)$ defines a continuous-time trajectory, and $t_i^{\mathrm{in}}$ and $t_i^{\mathrm{out}}$ are not restricted to multiples of $t_\mathrm{s}$. Let $t_i^{\mathrm{arr}}$ and $t_i^{\mathrm{exit}}$ denote the times at which vehicle $i$ enters and leaves a predefined coordination area, which determines which vehicles take part in the coordination at every time. At sampling instant $k$ the \emph{coordination set} is $\mathcal{C}_k \coloneqq \{\, i : t_i^{\mathrm{arr}}\le k t_\mathrm{s}\le t_i^{\mathrm{exit}} \,\}$, with cardinality $M\coloneqq|\mathcal{C}_k|$. The crossing order at sampling instant $k$ is a permutation $S_k$ of the elements of $\mathcal{C}_k$. Collision-free schedule is enforced through the precedence constraints
\begin{align*}
    t_{S_{c+1}}^{\mathrm{in}}\ge t_{S_{c}}^{\mathrm{out}},
    \qquad c\in\mathbb{I}_1^{M-1}.
\end{align*}
Collecting the decision variables in $x\coloneqq(x_i)_{i\in\mathcal{C}_k}$, $u\coloneqq(u_i)_{i\in\mathcal{C}_k}$ and $t\coloneqq(t_i)_{i\in\mathcal{C}_k}$, the cost to be minimized is the sum of the individual vehicle costs, i.e.,
\begin{align}
    J(x,u) &\coloneqq \sum_{i\in\mathcal{C}_k} J_i(x_i,u_i), \label{eq:cost}\\
    J_i(x_i,u_i) &\coloneqq \sum_{n=0}^{N-1}
		\ell_i(x_{i,n}, u_{i,n}) + \ell_{i,\mathrm{f}}(x_{i,N}), \nonumber
\end{align}
with stage and terminal costs 
$\ell_i(x_{i,n},u_{i,n})=r_i u_{i,n}^2+q_i(v_{i,n}-v_{\mathrm{ref}})^2$ and $\ell_{i,\mathrm f}(x_{i,N})=q_i(v_{i,N}-v_{\mathrm{ref}})^2$, where $v_{i,n}$ is the velocity of vehicle $i$ at time step $n$, $v_{\mathrm{ref}}$ is the reference velocity, and the weights $q_i,r_i>0$ penalize the deviation from it and the control effort, respectively, and depend on the class of the vehicle (i.e., light, medium, or heavy).

\subsection{Centralized formulation}
\label{sec:monolithic}

For a fixed coordination set $\mathcal{C}_k$ and schedule $S_k$, the coordination problem of \cite{shahi2026lagrange,shahi2026mpc} reads
\begin{subequations}
\label{eq:monolithic}
\thickmuskip=3.5mu plus 2mu
\begin{align}
    V(S_k)\coloneqq\min_{x,u,t}\;&\; J(x,u) \label{eq:mono_obj}\\
    \text{s.t.}\;&\; x_{i,0}=\hat x_{i,0}, &&\!\!\!\!\forall i\in\mathcal{C}_k,\label{eq:mono_ic}\\
    &\; x_{i,n+1}=f_i(x_{i,n},u_{i,n}), &&\!\!\!\!\forall i\in\mathcal{C}_k,\ n\in\mathbb{I}_0^{N-1},\label{eq:mono_dyn}\\
    &\; h_i(x_{i,n},u_{i,n})\le 0, &&\!\!\!\!\forall i\in\mathcal{C}_k,\ n\in\mathbb{I}_0^{N-1},\label{eq:mono_path}\\
    &\; t_{S_{c+1}}^{\mathrm{in}}\ge t_{S_{c}}^{\mathrm{out}}, &&\!\!\!\!\forall c\in\mathbb{I}_1^{M-1},\label{eq:mono_prec}\\
    &\; p_i(t_i^{\dagger},x_i,u_i)=p_i^{\dagger}, &&\!\!\!\!\forall i\in\mathcal{C}_k,\ \dagger\in\{\mathrm{in},\mathrm{out}\},\label{eq:mono_cross}
\end{align}
\end{subequations}
where $\hat x_{i,0}$ is the measured state of vehicle $i$. Problem \eqref{eq:monolithic} is an NLP in the continuous variables $(x,u,t)$, solved for a \emph{fixed} schedule $S_k$. The joint optimization over $S_k$ and $(x,u,t)$ turns it into an MINLP, since the schedule is encoded by binary variables that select, for every pair of vehicles, which one crosses first \cite{hult_theory,fayazi}. Its combinatorial part is addressed, as in \cite{shahi2026lagrange,shahi2026mpc}, by a multiplier-guided swapping heuristic which we describe below.

\subsection{Multiplier-guided schedule refinement}
\label{sec:heuristic}

The combinatorial part of \eqref{eq:monolithic} is addressed in \cite{shahi2026lagrange,shahi2026mpc} by a local order swap over adjacent pairs between vehicles in the crossing order, guided by the multipliers of the precedence constraints. 
For a given schedule $S_k$, let $\lambda_c \ge 0$, $c \in \mathbb{I}_1^{M-1}$, denote the Lagrange multiplier of the precedence constraint \eqref{eq:mono_prec} between the adjacent vehicles $S_c$ and $S_{c+1}$, so that the Lagrangian of \eqref{eq:monolithic} contains the term $\lambda_c\,(t_{S_c}^{\mathrm{out}} - t_{S_{c+1}}^{\mathrm{in}})$. As recalled in \cite{shahi2026lagrange} and well-known from sensitivity analysis \cite{nocedal,fiacco}, $\lambda_c$ is the first-order sensitivity of the optimal cost with respect to a perturbation of this constraint: relaxing it by $\theta>0$, i.e., imposing $t_{S_{c+1}}^{\mathrm{in}}\ge t_{S_{c}}^{\mathrm{out}}-\theta$, decreases $V(S_k)$ by $\lambda_c\theta$, up to first order. A large $\lambda_c$ therefore marks the pair $(S_c,S_{c+1})$ as the most expensive one. 

The multiplier $\lambda_c$ predicts the cost change only for a small relaxation $\theta$ of the precedence constraint, whereas a swap chnages the order of the two vehicles, which is a finite change of the constraints of \eqref{eq:monolithic}. A large $\lambda_c$ therefore might indicate a promising swap but does not guarantee that the swap reduces the cost. Every candidate is hence verified: the schedule $S_{\mathrm{sw}} \coloneqq \mathrm{swap}(S_k,c)$ is evaluated by re-solving \eqref{eq:monolithic}, and the swap is accepted only if $V(S_{\mathrm{sw}}) < V(S_k)$.

Algorithm~\ref{alg:swap} is the mechanism of heuristic, in which the set $\mathcal{T}$ keeps track of pairs already evaluated and excludes them from the candidate set $\mathcal{P}$. Here $Z$ collects the primal-dual solution of \eqref{eq:monolithic}. In the receding-horizon setting of \cite{shahi2026mpc}, Algorithm~\ref{alg:swap} is invoked whenever a vehicle enters the coordination area, the new vehicle appended at the tail of $S_{k-1}$ and the solution warm-started from the previous one. Throughout this paper, we use the variant denoted H1 in \cite{shahi2026mpc} because H1 required fewer NLP solves than H2 in all simulations. Algorithm~\ref{alg:swap} gives a monotone improvement of the cost: line~\ref{line:accept} accepts a swap only when the cost decreases, so no swaps are accepted based on the magnitude of $\lambda_c$ alone.

\begin{algorithm}[t]
\caption{Multiplier-guided schedule refinement \cite{shahi2026lagrange,shahi2026mpc}}
\label{alg:swap}
\small
\begin{algorithmic}[1]
\State \textbf{Input:} $\mathcal{C}_k$, schedule $S$, states $\{\hat x_{i,0}\}$, memory $\mathcal{T}$
\State \textbf{Output:} $S^\star$, $Z^\star$, $\mathcal{T}$
\State $S^\star \gets S$
\State Solve \eqref{eq:monolithic} for $S^\star$, obtain $(V^\star, Z^\star, \lambda^\star)$
\While{\textbf{true}}
  \State $\mathcal{P} \gets \{\, c \in \mathbb{I}_1^{M-1} :
         \{S^\star_c, S^\star_{c+1}\} \notin \mathcal{T} \,\}$
  \If{$\mathcal{P} = \emptyset$} \textbf{break} \EndIf
  \State $c^\star \gets \arg\max_{c \in \mathcal{P}} \lambda^\star_c$
  \State $\mathcal{T} \gets \mathcal{T} \cup
         \{\{S^\star_{c^\star}, S^\star_{c^\star+1}\}\}$
  \State $S_{\mathrm{sw}} \gets \mathrm{swap}(S^\star, c^\star)$
  \State Solve \eqref{eq:monolithic} for $S_{\mathrm{sw}}$, obtain
         $(V_{\mathrm{sw}}, Z_{\mathrm{sw}}, \lambda_{\mathrm{sw}})$
  \label{line:resolve}
  \If{$V_{\mathrm{sw}} < V^\star$}
  \label{line:accept}
    \State $(S^\star, V^\star, Z^\star, \lambda^\star) \gets
           (S_{\mathrm{sw}}, V_{\mathrm{sw}}, Z_{\mathrm{sw}},
           \lambda_{\mathrm{sw}})$
  \EndIf
\EndWhile
\State \textbf{return} $(S^\star, Z^\star, \mathcal{T})$
\end{algorithmic}
\end{algorithm}

\section{Decentralized formulation}
\label{sec:decentralized}

We now discuss why the centralized problem~\eqref{eq:monolithic} is reformulated, by stating the three bottlenecks of Algorithm~\ref{alg:swap}. First, the structure of the problem is not exploited: the cost and the constraints are separable across vehicles, only coupling between the vehicles is precedence constraints, yet~\eqref{eq:monolithic} is solved as a single NLP.
Second, the computation is centralized: the trajectories of all vehicles are optimized centrally, which requires the full state information of every vehicle. Third, every candidate swap is evaluated at line~\ref{line:resolve} by a complete re-solve of~\eqref{eq:monolithic},
and the memory $\mathcal{T}$ bounds the number of such evaluations but not their cost. We address these bottlenecks by reformulating the problem in a bi-level form first proposed in~\cite{hult2020}, in which each vehicle solves its own optimal control problem and only its crossing times are exchanged with a coordinator, and by exploiting the sensitivities of the value functions to replace expensive re-solves at line~\ref{line:resolve} by the solution of a QP.

\subsection{Bi-level reformulation}
\label{sec:bilevel}

By the decomposition principle of \cite{hult2020}, we separate the trajectory optimization of each vehicle from the coordination decision. This is possible because the cost \eqref{eq:cost} is a sum of per-vehicle  and the only coupling between vehicles is the precedence constraint \eqref{eq:mono_prec}, which involves the crossing times $t$ not the trajectory. For fixed crossing times, the trajectory optimization of each vehicle is therefore an independent OCP, and problem \eqref{eq:monolithic} is the equivalent bi-level form described below.

\subsubsection{Lower level}
For a given pair $t_i=(t_i^{\mathrm{in}},t_i^{\mathrm{out}})$ and initial state $\hat x_{i,0}$, each vehicle solves its own OCP, whose optimal solution defines the value function:
\begin{subequations}
\label{eq:lower}
\begin{align}
    V_i(\hat x_{i,0},t_i)\coloneqq\min_{x_i,u_i}\;&\; J_i(x_i,u_i)\label{eq:lower_obj}\\
    \text{s.t.}\;&\; x_{i,0}=\hat x_{i,0},\label{eq:lower_ic}\\
    &\; x_{i,n+1}=f_i(x_{i,n},u_{i,n}),\quad n\in\mathbb{I}_0^{N-1},\label{eq:lower_dyn}\\
    &\; h_i(x_{i,n},u_{i,n})\le 0,\quad n\in\mathbb{I}_0^{N-1},\label{eq:lower_path}\\
    &\; p_i(t_i^{\dagger},x_i,u_i)=p_i^{\dagger},\quad \dagger\in\{\mathrm{in},\mathrm{out}\}.\label{eq:lower_cross}
\end{align}
\end{subequations}
The $M$ problems defined as per \eqref{eq:lower} are independent and can be solved in parallel, as $t_i$ is imposed by a coordinator which solves the upper-level problem defined later. We collect the decision variables of \eqref{eq:lower} in $w_i\coloneqq(x_i,u_i)$ and write $p_i(t,w_i)$ for $p_i(t,x_i,u_i)$. The multipliers of the crossing conditions \eqref{eq:lower_cross} are denoted by $\nu_i\coloneqq(\nu_i^{\mathrm{in}},\nu_i^{\mathrm{out}})$, and the optimal primal solution of \eqref{eq:lower} and these multipliers, as functions of $t_i$, are denoted respectively by $w_i(t_i)$ and $\nu_i(t_i)$.

Problem \eqref{eq:lower} is not feasible for every $t_i$, since the vehicle can reach $p_i^{\mathrm{in}}$ and $p_i^{\mathrm{out}}$ only within limited time intervals so the value function $V_i$ is therefore defined only on its domain, i.e., the set of $t_i$ for which \eqref{eq:lower} is feasible, and this domain must be imposed as a constraint of the upper level introduced below. Following \cite{hult_primal,hult2020}, we impose it through the conditions $d_i(\hat x_{i,0},t_i)\ge0$, i.e.,
\begin{subequations}
\label{eq:domain}
\begin{align}
    \underline t_i^{\mathrm{in}}&\le t_i^{\mathrm{in}}\le \bar t_i^{\mathrm{in}},\\
    \underline t_i^{\mathrm{out}}(t_i^{\mathrm{in}})&\le t_i^{\mathrm{out}}\le \bar t_i^{\mathrm{out}}(t_i^{\mathrm{in}}).\label{eq:domain_out}
\end{align}
\end{subequations}
The earliest entry time $\underline t_i^{\mathrm{in}}$ is reached from $\hat x_{i,0}$ with maximal acceleration, and the latest entry time $\bar t_i^{\mathrm{in}}$ is the last time at which the vehicle can still brake without passing $p_i^{\mathrm{in}}$. The exit-time bounds $\underline t_i^{\mathrm{out}}$ and $\bar t_i^{\mathrm{out}}$ follow from entering at $t_i^{\mathrm{in}}$ with the highest or the lowest reachable speed and then accelerating or braking maximally. All times are also bounded by the end $(k+N)t_\mathrm{s}$ of the prediction horizon. For the double integrator of Section~\ref{sec:results}, all bounds are given in closed form by profiles with at most one switch between maximal acceleration and deceleration. Since the entry speeds depend on $t_i^{\mathrm{in}}$, the bounds on the exit time are functions of the entry time, and $d_i$ is nonlinear.

\subsubsection{Upper level}
\begin{figure}[!t]
\centering
\begin{tikzpicture}[
  font=\footnotesize, >={Latex[length=1.6mm]},
  box/.style={draw, rounded corners=2pt, align=center, inner sep=3pt},
  veh/.style={box, minimum width=1.55cm, minimum height=1.0cm},
  lab/.style={font=\scriptsize, inner sep=1pt, align=center}]
\node[veh] (v1) at (0,0) {\textbf{Vehicle $1$}\\ OCP \eqref{eq:lower}\\ \texttt{acados}};
\node (d1) at (1.45,0) {$\cdots$};
\node[veh] (vi) at (2.9,0) {\textbf{Vehicle $i$}\\ OCP \eqref{eq:lower}\\ \texttt{acados}};
\node (d2) at (4.35,0) {$\cdots$};
\node[veh] (vM) at (5.8,0) {\textbf{Vehicle $M$}\\ OCP \eqref{eq:lower}\\ \texttt{acados}};
\node[box, minimum width=7.35cm, minimum height=1.2cm, anchor=south] (up) at (2.9,1.75) {%
  \textbf{Coordinator (upper level)}\\[1pt]
  SQP on $t$: QP \eqref{eq:qp} with \texttt{HPIPM} (dense / structured)\\
  swap prediction: prediction QP \eqref{eq:surrogate}, ranking by $\lambda_c$};
\foreach \v/\j in {v1/1,vi/i,vM/M}{
  \draw[->, thick] ([xshift=-0.42cm]\v.north |- up.south) -- node[lab, left]{$t_{\j}$} ([xshift=-0.42cm]\v.north);
  \draw[->, thick] ([xshift=0.42cm]\v.north) -- node[lab, right]{$V_{\j}$, $d_{\j}$\\and their\\derivatives} ([xshift=0.42cm]\v.north |- up.south);
}
\end{tikzpicture}
\caption{Bi-level architecture. The coordinator sends the crossing times $t_i$ (downward arrows) to each vehicle, which solves its own OCP and returns its value function $V_i$ and domain constraints $d_i$ with their gradients and Hessians (upward arrows).}
\label{fig:architecture}
\end{figure}
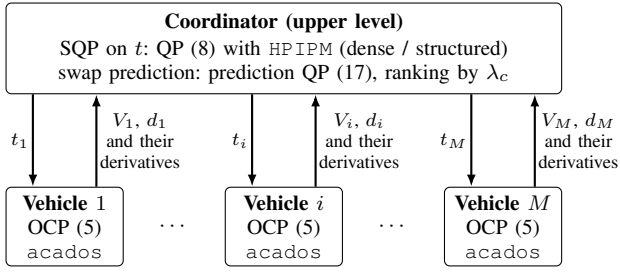

The upper level is a coordinator which can solve the overall coordination problem by accessing only the value functions $V_i$, the constraints $d_i$, and their derivatives, without exchanging the trajectories. The crossing times and the schedule are then optimized by solving~\cite{hult2020}
\begin{subequations}
\label{eq:upper}
\begin{align}
    \min_{S,\,t}\;&\; \bar V(S,t)\coloneqq\sum_{i\in\mathcal{C}_k} V_i(\hat x_{i,0},t_i)\label{eq:upper_obj}\\
    \text{s.t.}\;&\; t_{S_{c+1}}^{\mathrm{in}}\ge t_{S_{c}}^{\mathrm{out}},\quad c\in\mathbb{I}_1^{M-1},\label{eq:upper_prec}\\
    &\; d_i(\hat x_{i,0},t_i)\ge 0,\quad i\in\mathcal{C}_k.\label{eq:upper_dom}
\end{align}
\end{subequations}
For a fixed $S_k$, \eqref{eq:upper} is a smooth NLP in $t$ only, and its optimal value coincides with the cost of the centralized problem,
\begin{align*}
    V(S_k)=\min_{t}~\bar V(S_k,t)\text{ s.t. } \eqref{eq:upper_prec},~\eqref{eq:upper_dom}.
\end{align*}
By construction, \eqref{eq:upper} is an exact reformulation of \eqref{eq:monolithic}: the lower level optimizes the trajectory variables and exposes only the value functions $V_i$ to the upper level, which retains the precedence coupling on the $2M$ crossing times. The multipliers $\lambda_c$ of the precedence constraints \eqref{eq:upper_prec} are those used to rank the candidate swaps in Algorithm~\ref{alg:swap}, and they are now available directly from the upper-level solution. Fig.~\ref{fig:architecture} illustrates the resulting architecture. The coordinator sends the crossing times $t_i$ to each vehicle, each vehicle solves its OCP~\eqref{eq:lower} independently and returns the value $V_i$ and the constraints $d_i$ together with their gradients and Hessians, and the coordinator uses this information to update the crossing times and the schedule. No trajectory information is exchanged between the two levels.

\subsection{Upper-level SQP}
\label{sec:sqp}
For a fixed schedule $S$, the upper-level NLP \eqref{eq:upper} is solved at every replanning and again for every confirmed candidate swap. Its computation time therefore determines that of the coordinator. We solve it by SQP \cite{nocedal}, following \cite{hult2020}, with a custom implementation in Python in which the lower-level OCPs \eqref{eq:lower} are solved as subproblems. Since SQP uses the Hessian of the Lagrangian, it needs only a few iterations when it is warm-started from the previous solution. The quadratic model built at the solution is also used in Section~\ref{sec:swap} to predict the effect of a swap. The method generates a sequence $t^{(0)},t^{(1)},\ldots$ of crossing-time vectors, starting from a warm start $t^{(0)}$. Each iterate $t^{(j)}$ is a tentative assignment of the entry and exit times of all vehicles in $\mathcal{C}_k$, at which the vehicles solve \eqref{eq:lower} and return $V_i$ and $d_i$ together with their gradients and Hessians with respect to $t_i$, computed as described in Sections~\ref{sec:derivatives} and~\ref{sec:sensitivity}, and the sequence converges to the optimal crossing times for $S$. The iterate $t^{(j)}$ is updated with the step $\Delta t^{(j)}$ that solves the QP:
\begin{subequations}
\label{eq:qp}
\begin{align}
    \min_{\Delta t}\;&\; \tfrac12\,\Delta t^{\top}\bar H^{(j)}\Delta t
        +\nabla_t \bar V\big(S,t^{(j)}\big)^{\top}\Delta t \label{eq:qp_obj}\\
    \text{s.t.}\;&\; t_{S_{c+1}}^{\mathrm{in},(j)}+\Delta t_{S_{c+1}}^{\mathrm{in}}\ge t_{S_c}^{\mathrm{out},(j)}+\Delta t_{S_c}^{\mathrm{out}},\quad c\in\mathbb{I}_1^{M-1},\label{eq:qp_prec}\\
    &\; d_i\big(\hat x_{i,0},t_i^{(j)}\big)+\nabla_{t_i} d_i\big(\hat x_{i,0},t_i^{(j)}\big)^{\top}\Delta t_i\ge 0,\quad i\in\mathcal{C}_k,
    \label{eq:qp_dom}
\end{align}
\end{subequations}
in which the precedence constraints are kept exactly, and the domain constraints are linearized using the gradient. We denote by $\lambda$ and $\mu_i$ the multipliers of \eqref{eq:qp_prec} and \eqref{eq:qp_dom}, respectively. To guarantee convexity of \eqref{eq:qp}, each block $H_i$ of the Hessian of the Lagrangian \eqref{eq:blockH} is made positive definite by the eigenvalue modification of \cite{hult2020}: with $H_i=E_iD_iE_i^{\top}$ its eigendecomposition, we set $\bar H_i=E_i\bar D_iE_i^{\top}$, where the diagonal matrix $\bar D_i$ is obtained from $D_i$ by replacing every eigenvalue smaller than a small $\epsilon>0$ by $\epsilon$, and take $\bar H^{(j)}=\operatorname{diag}(\bar H_i)_{i\in\mathcal{C}_k}$, evaluated at $t^{(j)}$. Note that these decompositions are cheap because the blocks have dimension $2$. To save computations, the blocks $\nabla^2_{t_i}V_i$ are recomputed only every second iteration and reused in between. 

Far from the solution, the full step $\Delta t^{(j)}$ may increase the cost, and SQP is then not guaranteed to converge. We therefore shorten the step by a customized backtracking line search, in which every trial point is first mapped by an operator $\bar \Pi$ into the set of feasible exit times, given the current entry times, as otherwise the cost cannot be evaluated. The projection then acts on $t_i^\mathrm{out}$ and reads
\begin{align*}
    \bar \Pi(t_i^{\mathrm{out},j})&\coloneqq \min\left(\max\left(t_i^\mathrm{out},\,\underline t_i^{\mathrm{out}}(\hat t)\right),\, \bar t_i^{\mathrm{out}}(\hat t)\right), \\
    \hat t &\coloneqq t_i^{\mathrm{in},j}+\alpha^{(j)} \Delta t_i^{\mathrm{in},j},
\end{align*}
where $\alpha^{(j)}$ is the step size.

This projection can be computed separately for each vehicle, and is cheap because the constraint needs anyway to be evaluated for the updated $t_i^{\mathrm{in}}$. We denote as $\Pi$ the projection operator applied to $t$, which leaves the entry times unchanged and applies $\bar \Pi$ to the exit times. 
 
The step size $\alpha^{(j)}\in(0,1]$ is reduced by a constant factor until the Armijo condition
\begin{multline*}
    \bar V\big(S,\Pi(t^{(j)}+\alpha^{(j)}\Delta t^{(j)})\big)\\
    \le \bar V\big(S,t^{(j)}\big)+\gamma\,\alpha^{(j)}\,\nabla_t\bar V\big(S,t^{(j)}\big)^{\top}\Delta t^{(j)}
\end{multline*}
holds, where $\gamma\in(0,\tfrac12]$ \cite{nocedal}. The iterative scheme then reads as:
\begin{equation}
    t^{(j+1)}=\Pi\big(t^{(j)}+\alpha^{(j)}\Delta t^{(j)}\big),\qquad
    (\lambda,\mu)^{(j+1)}=(\lambda,\mu)_{\mathrm{QP}}^{(j)},
\label{eq:sqp_update}
\end{equation}
where $(\lambda,\mu)_{\mathrm{QP}}^{(j)}$ are the multipliers of \eqref{eq:qp_prec} and \eqref{eq:qp_dom}. The iteration stops when $t^{(j)}$ satisfies \eqref{eq:upper_prec} and \eqref{eq:upper_dom} up to a numerical tolerance of $10^{-3}$ and, in addition, either $\|\Delta t^{(j)}\|_\infty\le10^{-3}$, i.e., no crossing time changes by more than \SI{1}{ms}, or the decrease predicted by the QP model is below $10^{-5}\max(1,|\bar V|)$. 
The multipliers $\lambda_c$ used in Algorithm~\ref{alg:swap} are those of \eqref{eq:qp_prec} at the last iterate. Each replanning starts from the previous crossing times, with those of a new vehicle appended at the tail of the schedule, and each vehicle warm-starts its OCP from its own previous primal-dual solution. The multipliers of the upper level are initialized at zero. Since consecutive coordination sets differ by at most one vehicle, only a few SQP iterations are needed.

The QP \eqref{eq:qp} is solved with \texttt{HPIPM} \cite{hpipm} in two different formulations. The \emph{dense} formulation passes the $2M$ crossing times, the block-diagonal Hessian $\bar H^{(j)}$ and the constraint Jacobian as dense matrices. The \emph{structured} formulation orders the variables along the schedule, to generate the fictitious dynamical system 
\begin{align}
    t_{S_{i+1}}= t_{S_{i}} + \delta t_{S_{i}},
    \label{eq:time_dynamics}
\end{align}
with the entry and exit times $t_{S_{i}}$ as states and $\delta t_{S_c}$ as control inputs. The cost \eqref{eq:qp_obj} is unchanged and splits into one term per stage, because $\bar H^{(j)}$ is block diagonal, and the constraints are
\begin{align*}
    t_{S_{c+1}}^{\mathrm{in},(j)}+\Delta t_{S_{c+1}}^{\mathrm{in}}-t_{S_c}^{\mathrm{out},(j)}-\Delta t_{S_c}^{\mathrm{out}}&\ge0,\quad c\in\mathbb{I}_1^{M-1},\\
    d_{S_c}\big(t_{S_c}^{(j)}\big)+\nabla_{t_{S_c}}d_{S_c}\big(t_{S_c}^{(j)}\big)^{\top}\Delta t_{S_c}&\ge0,\quad c\in\mathbb{I}_1^{M}.
\end{align*}
As it is written, the first constraint is not local in time. However, we can exploit~\eqref{eq:time_dynamics} to write
\begin{align*}
t_{S_{c+1}}^{\mathrm{in},(j)}+\Delta t_{S_{c+1}}^{\mathrm{in}} &= t_{S_{c}}^{\mathrm{in},(j)}+\Delta t_{S_{c}}^{\mathrm{in}} + \delta t_{S_{c}}^{\mathrm{in},(j)}+\Delta \delta t_{S_{c}}^{\mathrm{in}},
\end{align*}
which allows us to rewrite the constraint such that it is now local in time.
Each precedence constraint couples only two consecutive stages, so \eqref{eq:qp} is an OCP and \texttt{HPIPM} solves it by a Riccati recursion whose cost grows linearly with $M$. The dense formulation is faster for few vehicles and the structured one for many, so we choose between them by comparing $M$ with the crossover size $\bar M$ found in Section~\ref{sec:res_qp}.

\subsection{Derivatives of the value function and of the domain constraints}
\label{sec:derivatives}
The QP \eqref{eq:qp} requires the first and second derivatives of $V_i$ and $d_i$ with respect to $t_i$. Those of $V_i$ are obtained from the KKT system of the lower-level OCP, and those of $d_i$ from its closed form.

\subsubsection{First-order derivatives}
The parameter $t_i$ enters \eqref{eq:lower} only through the crossing conditions \eqref{eq:lower_cross}, which appear in the Lagrangian function of \eqref{eq:lower} as the terms $\nu_i^{\dagger}\big(p_i(t_i^{\dagger},w_i)-p_i^{\dagger}\big)$. By the implicit function theorem applied to the KKT conditions of \eqref{eq:lower}, under the assumptions of Section~\ref{sec:sensitivity}, the derivative of $V_i$ with respect to $t_i^{\dagger}$ equals the partial derivative of these terms with respect to $t_i^{\dagger}$, evaluated at the primal-dual solution \cite{fiacco,nocedal}:
\begin{equation}
    \frac{\partial V_i}{\partial t_i^{\dagger}}
    =\nu_i^{\dagger}(t_i)\,\frac{\partial p_i}{\partial t}\big(t_i^{\dagger},w_i(t_i)\big),
    \label{eq:grad_component}
\end{equation}
where $\partial p_i/\partial t$ is the velocity of vehicle $i$ at the crossing time $t_i^{\dagger}$. The two components \eqref{eq:grad_component} form $\nabla_{t_i}V_i$, and the gradient of the upper-level cost \eqref{eq:upper_obj} is $\nabla_t \bar V=\big(\nabla_{t_i} V_i\big)_{i\in\mathcal{C}_k}$.

\subsubsection{Second-order derivatives}
Differentiating \eqref{eq:grad_component} with respect to $t_i^{\ddagger}$ by the product and chain rules \cite{fiacco,buskens_maurer} yields
\begin{multline}
    \frac{\partial^2 V_i}{\partial t_i^{\dagger}\,\partial t_i^{\ddagger}}
    =\frac{\mathrm d\nu_i^{\dagger}}{\mathrm dt_i^{\ddagger}}\,\frac{\partial p_i}{\partial t}
     +\nu_i^{\dagger}\,\frac{\partial^2 p_i}{\partial t\,\partial w_i}\,
      \frac{\mathrm dw_i}{\mathrm dt_i^{\ddagger}}\\
     +\begin{cases}\nu_i^{\dagger}\,\dfrac{\partial^2 p_i}{\partial t^2}, & \ddagger=\dagger,\\[4pt] 0, & \ddagger\neq\dagger,\end{cases}
    \label{eq:hess_component}
\end{multline}
with all partial derivatives of $p_i$ evaluated at $\big(t_i^{\dagger},w_i(t_i)\big)$. The second term accounts for the change of the optimal trajectory, and the last term, the acceleration at $t_i^{\dagger}$ weighted by the multiplier, appears only on the diagonal, because $t_i^{\dagger}$ is the first argument of $p_i$ in \eqref{eq:grad_component}. The sensitivities $\mathrm d\nu_i/\mathrm dt_i$ and $\mathrm dw_i/\mathrm dt_i$ are computed in Section~\ref{sec:sensitivity}. Since the Hessian $\nabla^2_{t_i}V_i$ is symmetric in exact arithmetic, we compute only its diagonal entries and the entry with $\dagger=\mathrm{in}$, $\ddagger=\mathrm{out}$ from \eqref{eq:hess_component}, and use the latter for both off-diagonal entries.

\subsubsection{Derivatives of the domain constraints}
The constraints $d_i$ depend only on $t_i$ and on the measured state $\hat x_{i,0}$, not on the solution of \eqref{eq:lower}. By \eqref{eq:domain}, their components are
\begin{align*}
    d_i(\hat x_{i,0},t_i)=\begin{bmatrix}
        t_i^{\mathrm{in}}-\underline t_i^{\mathrm{in}}\\
        \bar t_i^{\mathrm{in}}-t_i^{\mathrm{in}}\\
        t_i^{\mathrm{out}}-\underline t_i^{\mathrm{out}}(t_i^{\mathrm{in}})\\
        \bar t_i^{\mathrm{out}}(t_i^{\mathrm{in}})-t_i^{\mathrm{out}}
    \end{bmatrix},
\end{align*}
where $\underline t_i^{\mathrm{in}}$ and $\bar t_i^{\mathrm{in}}$ do not depend on $t_i$ and the constraints are linear in $t_i^{\mathrm{out}}$. The first two components are linear in $t_i^{\mathrm{in}}$, and only the exit time bounds contribute to the Lagrangian Hessian. Their gradients and Hessians with respect to $t_i$ are
\begin{align*}
    \nabla_{t_i}d_{i,3}&=\begin{bmatrix}-\underline t_i^{\mathrm{out}\,\prime}\\ 1\end{bmatrix},&
    \nabla^2_{t_i}d_{i,3}&=\begin{bmatrix}-\underline t_i^{\mathrm{out}\,\prime\prime}&0\\0&0\end{bmatrix},\\
    \nabla_{t_i}d_{i,4}&=\begin{bmatrix}\bar t_i^{\mathrm{out}\,\prime}\\ -1\end{bmatrix},&
    \nabla^2_{t_i}d_{i,4}&=\begin{bmatrix}\bar t_i^{\mathrm{out}\,\prime\prime}&0\\0&0\end{bmatrix},
\end{align*}
where $\prime$ and $\prime\prime$ denote the first and second derivatives of the scalar function. Under mild assumptions, we can define $\bar t_i^{\mathrm{out}}(t_i^{\mathrm{in}})$ and $\underline t_i^{\mathrm{out}}(t_i^{\mathrm{in}})$ through the implicit functions
\begin{subequations}
\begin{align}
    p_i(\bar t_i^{\mathrm{out}}(t_i^{\mathrm{in}}),\bar x_i(t_i^{\mathrm{in}}),\bar u_i(t_i^{\mathrm{in}}))=p_i^{\mathrm{out}}, \label{eq:t_out_ub}\\
    p_i(\underline t_i^{\mathrm{out}}(t_i^{\mathrm{in}}),\underline x_i(t_i^{\mathrm{in}}),\underline u_i(t_i^{\mathrm{in}}))=p_i^{\mathrm{out}}, \label{eq:t_out_lb}
\end{align}
\label{eq:t_out_bounds}
\end{subequations}
where
\begin{subequations}
\begin{align}
    (\bar x_i(t_i^{\mathrm{in}}),\bar u_i(t_i^{\mathrm{in}}))\coloneqq\arg\min_{x_i,u_i}\;&\; p_{i,N}\\
    \text{s.t.}\;&\; \eqref{eq:lower_ic},\eqref{eq:lower_dyn},\eqref{eq:lower_path}\\
    &\; p_i(t_i^{\mathrm{in}},x_i,u_i)=p_i^{\mathrm{in}},
\end{align}
\label{eq:nlp_bounds}
\end{subequations}
and $\underline x_i(t_i^{\mathrm{in}}),\underline u_i(t_i^{\mathrm{in}})$ are obtained by solving the same problem by flipping the sign of the cost. The derivatives of $\bar t_i^{\mathrm{out}}(t_i^{\mathrm{in}})$ and $\underline t_i^{\mathrm{out}}(t_i^{\mathrm{in}})$ are obtained by algorithmic differentiation applied to~\eqref{eq:t_out_bounds}. In case the dynamics are simple enough, one can obtain an analytical solution to~\eqref{eq:nlp_bounds}, e.g., always apply maximum braking or maximum acceleration. In the general case, the problem can be solved numerically, but the complexity of evaluating and differentiating the solution is essentially the same as the complexity of evaluating and differentiating the cost.

The Hessian of the Lagrangian of \eqref{eq:upper} with respect to $t$ contains the Hessian of the cost \eqref{eq:upper_obj} and the second derivatives of the constraints, weighted by their multipliers. The precedence constraints \eqref{eq:upper_prec} are affine and do not contribute, while the domain constraints \eqref{eq:upper_dom} do. Since the cost is separable and each $d_i$ depends only on $t_i$, this Hessian is block diagonal with one $2\times2$ block per vehicle,
\begin{equation}
    H_i\coloneqq\nabla^2_{t_i}V_i-\sum_{m}\mu_{i,m}\,\nabla^2_{t_i}d_{i,m},
    \label{eq:blockH}
\end{equation}
where $\mu_{i,m}\ge0$ is the $m$-th component of $\mu_i$, taken from the previous QP. Only the $(\mathrm{in},\mathrm{in})$ entry of the second term is nonzero, and it equals $\mu_{i,3}\,\underline t_i^{\mathrm{out}\,\prime\prime}-\mu_{i,4}\,\bar t_i^{\mathrm{out}\,\prime\prime}$.

\subsection{Parametric sensitivity analysis}
\label{sec:sensitivity}
The expression \eqref{eq:hess_component} requires the sensitivities $\mathrm dw_i/\mathrm dt_i$ and $\mathrm d\nu_i/\mathrm dt_i$ of the lower-level primal-dual solution. We assume that the solution of \eqref{eq:lower} satisfies the linear independence constraint qualification, strict complementarity and the second-order sufficient conditions \cite{fiacco,nocedal}. Then, by the implicit function theorem applied to the KKT conditions of \eqref{eq:lower}, the primal-dual solution is locally a differentiable function of $t_i$, and its derivative is the solution of a linear system whose matrix is the KKT matrix at the solution and whose right-hand side is the derivative of the KKT residual with respect to $t_i$ \cite{fiacco,buskens_maurer,frey2025differentiable}. Since $t_i$ has two components, two such systems are solved, and their solutions provide $\mathrm dw_i/\mathrm dt_i$ and $\mathrm d\nu_i/\mathrm dt_i$ for \eqref{eq:hess_component}.

We compute these sensitivities with the solution-sensitivity functionality of \texttt{acados} \cite{acados,frey2025differentiable}. The crossing conditions \eqref{eq:lower_cross} are implemented by augmenting the state with the positions of the vehicle at $t_i^{\mathrm{in}}$ and $t_i^{\mathrm{out}}$, which are recorded within the sampling interval that contains the respective crossing time and constrained at the final stage, so that $\nu_i$ are the multipliers of these terminal constraints. The OCP is solved by SQP with the exact Hessian of the Lagrangian, which is required for exact sensitivities \cite{frey2025differentiable}, and its QP subproblems are solved by \texttt{HPIPM} \cite{hpipm} with partial condensing \cite{frison_condensing}. At the converged solution, the KKT matrix of the interior-point formulation, which contains all inequality constraints, is formed with the exact Hessian and factorized once by the classical Riccati recursion, which requires positive definiteness only of the reduced Hessian, as ensured by the second-order sufficient conditions \cite{frey2025differentiable}. The forward sensitivities with respect to the two crossing times are then obtained by two back-substitutions with this factorization and are evaluated only at the stages that contain $t_i^{\mathrm{in}}$ and $t_i^{\mathrm{out}}$ and at the final stage.

\subsection{Sensitivity-based swap prediction}
\label{sec:swap}

Line~\ref{line:resolve} of Algorithm~\ref{alg:swap} evaluates a candidate swap by re-solving the coordination problem for the swapped schedule. In the bi-level form, each such evaluation costs one complete upper-level SQP solve together with the associated lower-level OCPs. We discuss next how the derivatives introduced in Section~\ref{sec:derivatives} can replace most of these re-solves by a single QP.

Let $t^{\star}=t^{\star}(S)$ be the converged upper-level solution for the schedule $S$, let $\nabla_t \bar V\coloneqq\nabla_t\bar V(S,t^\star)$ be the gradient and $\bar H$ the Hessian of the Lagrangian \eqref{eq:blockH} after the eigenvalue modification of Section~\ref{sec:sqp}, both evaluated at $t^\star$, and let $\lambda$ be the multipliers of \eqref{eq:upper_prec}. Around $t^{\star}$, the cost admits a second-order approximation, in which a small step $\delta t$ is distinguished from the SQP step $\Delta t$:
\begin{equation}
    \bar V(S,t^{\star}+\delta t)\approx \bar V(S,t^{\star})
    +\nabla_t \bar V^{\top}\delta t+\tfrac12\,\delta t^{\top}\bar H\,\delta t .
    \label{eq:quadmodel}
\end{equation}
A swap of the adjacent pair $c$ replaces the precedence constraints \eqref{eq:upper_prec} by those of $S_{\mathrm{sw}}=\mathrm{swap}(S,c)$. The effect of the swap is predicted by minimizing the model \eqref{eq:quadmodel} subject to the linearized constraints of $S_{\mathrm{sw}}$, i.e., by the prediction QP:
\begin{subequations}
\label{eq:surrogate}
\begin{align}
    \Delta J_{\mathrm{pred}}(c)\coloneqq\min_{\delta t}\;&\;
        \nabla_t \bar V^{\top}\delta t+\tfrac12\,\delta t^{\top}\bar H\,\delta t \label{eq:surr_obj}\\
    \text{s.t.}\;&\; \eqref{eq:qp_prec}\text{ for }S_{\mathrm{sw}}\text{ and }\eqref{eq:qp_dom},\\ &\;\text{with } t^{(j)}=t^\star\text{ and }\Delta t=\delta t.
    \label{eq:surr_con}
\end{align}
\end{subequations}
Since $t^\star$ is a KKT point for $S$, the same QP with the constraints of $S$ has zero optimal value, so $\Delta J_{\mathrm{pred}}(c)$ approximates the cost change $V(S_{\mathrm{sw}})-V(S)$. Problem \eqref{eq:surrogate} has the same structure as \eqref{eq:qp} and is solved with the same \texttt{HPIPM} formulation. It involves no lower-level solve, because it reuses the gradient and the Hessian already computed at $t^{\star}$.

The pair $c$ is predicted to improve the cost when $\Delta J_{\mathrm{pred}}(c)<0$. Because the swap is a discrete, finite perturbation of the schedule, the prediction is used as a screen. Pairs with $\Delta J_{\mathrm{pred}}(c)\ge 0$ are discarded at the cost of one QP, whereas a pair predicted to improve is confirmed by a warm-started re-solve of \eqref{eq:upper} for $S_{\mathrm{sw}}$ and accepted only if $V(S_{\mathrm{sw}})<V(S)$. Algorithm~\ref{alg:predict} reports the resulting procedure. It preserves the acceptance rule and hence the monotonicity of Algorithm~\ref{alg:swap}, uses the same memory $\mathcal{T}$, and replaces most of the re-solves by QPs.

\begin{algorithm}[t]
\caption{Sensitivity-based schedule refinement}
\label{alg:predict}
\small
\begin{algorithmic}[1]
\State \textbf{Input:} $\mathcal{C}_k$, schedule $S$, states $\{\hat x_{i,0}\}$, warm start $t$, memory $\mathcal{T}$
\State \textbf{Output:} $S^\star$, $t^\star$, $\mathcal{T}$
\State $S^\star \gets S$
\State Solve \eqref{eq:upper} for $S^\star$ to obtain $\mathcal{D}^\star\coloneqq(V^\star, t^\star, \lambda^\star, \nabla_t\bar V, \bar H)$
\While{\textbf{true}}
  \State $\mathcal{P} \gets \{\, c \in \mathbb{I}_1^{M-1} :
         \{S^\star_c, S^\star_{c+1}\} \notin \mathcal{T} \,\}$
  \If{$\mathcal{P} = \emptyset$} \textbf{break} \EndIf
  \State $c^\star \gets \arg\max_{c \in \mathcal{P}} \lambda^\star_c$
  \State $\mathcal{T} \gets \mathcal{T} \cup
         \{\{S^\star_{c^\star}, S^\star_{c^\star+1}\}\}$
  \State Solve \eqref{eq:surrogate} for $c^\star$, obtain $\Delta J_{\mathrm{pred}}(c^\star)$ and $\delta t$
  \If{$\Delta J_{\mathrm{pred}}(c^\star) < 0$}
    \State $S_{\mathrm{sw}} \gets \mathrm{swap}(S^\star, c^\star)$
    \State Obtain $\mathcal{D}_{\mathrm{sw}}\coloneqq(V_{\mathrm{sw}}, t_{\mathrm{sw}}, \lambda_{\mathrm{sw}}, \nabla_t\bar V_{\mathrm{sw}}, \bar H_{\mathrm{sw}})$ by solving \eqref{eq:upper} for $S_{\mathrm{sw}}$
    \If{$V_{\mathrm{sw}} < V^\star$}
      \State $\mathcal{D}^\star \gets\mathcal{D}_{\mathrm{sw}}$
    \EndIf
  \EndIf
\EndWhile
\State \textbf{return} $(S^\star, t^\star, \mathcal{T})$
\end{algorithmic}
\end{algorithm}

\subsection{Real-time scheduling iteration}
\label{sec:realtime}

Algorithm~\ref{alg:predict}, though computationally cheaper than Algorithm~\ref{alg:swap}, still relies on complete SQP solves: one for the replanning and one for every confirmation. For closed-loop simulation with a fixed sampling time we also consider a real-time scheduling variant based on the idea of the real-time iteration \cite{rti}. At every sampling instant, the derivatives \eqref{eq:grad_component} and \eqref{eq:hess_component} are evaluated at the current lower-level solutions, which are available from the vehicles' own MPC controllers. Only a single full SQP step \eqref{eq:qp} is taken for the current schedule, and one prediction QP \eqref{eq:surrogate} is solved for the most promising pair not in $\mathcal{T}$. If this pair is predicted to improve, the swapped schedule is accepted together with the step of its prediction QP. A swap is accepted based on the prediction alone, and the confirmation of Algorithm~\ref{alg:predict} is replaced by the next sampling instant, which linearizes the problem again at the new schedule and updates the multiplier ranking. The computation per sampling instant consists of one derivative evaluation and two QPs, independently of the number of candidate swaps, so its computation time varies little between sampling instants.

\section{Numerical results}
\label{sec:results}

As in \cite{shahi2026lagrange,shahi2026mpc}, the dynamics \eqref{eq:dyn} are a double integrator, whose input $u_{i,n}$ is the acceleration. The path constraints \eqref{eq:path} impose $|u_{i,n}|\le\SI{2}{m/s^2}$ and $v_{i,n}\ge0$. All experiments use $v_{\mathrm{ref}}=\SI{13.89}{m/s}$ (\SI{50}{km/h}) and $t_\mathrm{s}=\SI{0.1}{s}$, and three vehicle classes, light, medium and heavy, with weights $(q_i,r_i)=(1,1)$, $(10,10)$ and $(100,500)$. The lower-level OCPs \eqref{eq:lower} are generated with \texttt{acados} and solved with \texttt{HPIPM}. In the architecture of Fig.~\ref{fig:architecture}, each vehicle solves its own OCP. In our simulations, the OCPs are solved on one computer in parallel, with one thread per vehicle. All computations were carried out on a laptop with an Intel Core i7-13620H processor (10 cores, 16 hardware threads, up to \SI{4.9}{GHz}).

\subsection{Distributed vs Centralized Solution}
\label{sec:res_random}

We compare the bi-level scheme with the centralized formulation \eqref{eq:monolithic} on $100$ randomized scenarios with six vehicles each. In every scenario, the vehicles start \SI{100}{m} upstream of the intersection at $v_{\mathrm{ref}}$, with inter-arrival times drawn uniformly from $[0.2,0.4]\,\si{s}$, and the horizon is $N=150$. The vehicle classes are drawn independently with probabilities $(0.4,0.2,0.4)$ for light, medium and heavy vehicles in $25$ scenarios, $(0.6,0.2,0.2)$ in $25$, $(0.2,0.2,0.6)$ in $25$ and uniformly in the remaining $25$. The centralized problem \eqref{eq:monolithic} includes the same domain constraints \eqref{eq:domain}. It is solved as one \texttt{acados} OCP that contains all vehicles, by SQP with a Gauss--Newton Hessian, and its schedule is refined by Algorithm~\ref{alg:swap}. Every arrival triggers both algorithms, with the new vehicle appended at the tail of the schedule. Both start each event from the same vehicle states, the same schedule and the same memory $\mathcal{T}$. The bi-level scheme uses Algorithm~\ref{alg:predict}. The bi-level scheme returns the same final crossing order as the centralized formulation in all $100$ scenarios.

The prediction QP was solved for $728$ candidate swaps. To assess it, every candidate that was discarded was also re-solved, outside the timing, which gives the true cost change $\Delta J_{\mathrm{true}}$. The sign of $\Delta J_{\mathrm{pred}}$ is correct for all of the $728$ candidates. All $228$ candidates with $\Delta J_{\mathrm{pred}}<0$ improve the cost and were accepted, so no re-solve was spent on a swap that does not improve. The relative error $|\Delta J_{\mathrm{true}}-\Delta J_{\mathrm{pred}}|/|\Delta J_{\mathrm{true}}|$ has a median of $3.4\%$, and large relative errors occur only for small $|\Delta J_{\mathrm{true}}|$.

\begin{figure}[!t]
\centering
\includegraphics[width=\columnwidth]{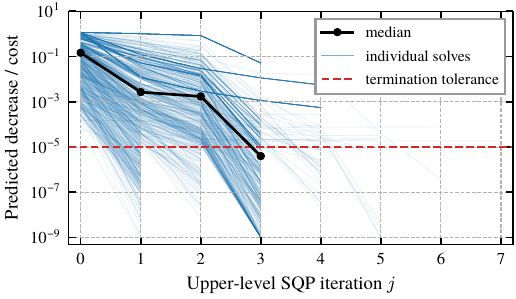}
\caption{Convergence of the upper-level SQP in the $100$ scenarios. Dashed: termination tolerance.}
\label{fig:s100_convergence}
\end{figure}

Fig.~\ref{fig:s100_convergence} shows the convergence of the upper-level SQP for all replanning and confirmation solves. The replanning converges in a median of $4$ SQP iterations and at most $8$. A confirmation $t^\star+\delta t$, converges in a median of $2$ and at most $7$. The QPs require a median of $7$ and at most $13$ \texttt{HPIPM} iterations. The centralized SQP, which optimizes all trajectories jointly, requires a median of $15$ and up to $149$ iterations per solve. The median computation time per event is \SI{382}{ms} for the centralized formulation and \SI{10.6}{ms} for the bi-level scheme. The largest event takes \SI{6.1}{s} and \SI{79}{ms}, respectively.

\subsection{Computation time of the bi-level scheme}
\label{sec:res_six}

We break down the computation time of Algorithm~\ref{alg:predict} on Scenario~1 of \cite{shahi2026mpc}. In this scenario, a heavy, a light, two medium and two heavy vehicles arrive within \SI{1.7}{s}, each starting \SI{100}{m} upstream of the intersection, and the horizon is $N=150$. Since $M\le6$, the dense QP formulation is used. The replanning for a fixed schedule takes at most \SI{15.8}{ms}. A confirmation requires two SQP iterations and \SI{4.90}{ms} on average, whereas a prediction costs \SI{0.40}{ms}, about $12$ times less. The prediction discarded five candidates, which saved about \SI{24.5}{ms} of confirmations at a prediction cost of \SI{2.0}{ms}. The most expensive event, with two confirmed swaps, takes \SI{28.4}{ms}, and the whole scenario requires \SI{103}{ms} of coordination time. All values are medians over five repeated simulations.

\begin{figure}[!t]
\centering
\includegraphics[width=\columnwidth]{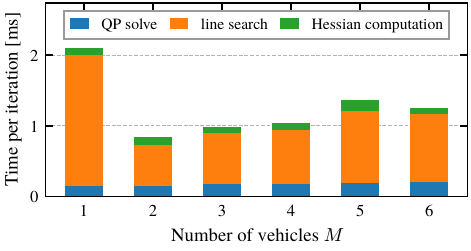}
\caption{Computation time per upper-level SQP iteration in the six-vehicle scenario versus the number of vehicles $M$.}
\label{fig:cost_upper}
\end{figure}

\begin{figure}[!t]
\centering
\includegraphics[width=\columnwidth]{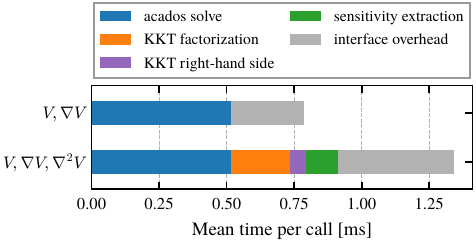}
\caption{Mean computation time per lower-level call in the six-vehicle scenario, with and without the Hessian.}
\label{fig:cost_lower}
\end{figure}

Fig.~\ref{fig:cost_upper} reports the computation time of one upper-level SQP iteration, split into its operations. A dense \texttt{HPIPM} QP takes \SI{0.18}{ms} on average with $7.5$ interior-point iterations. Most of the time is spent in the line search, because every trial point requires all vehicles to solve their OCPs to obtain $V_i$ and $\nabla_{t_i}V_i$. The accepted trial point provides these values for the next iterate, so this is the only lower-level solve per iteration. With one thread per vehicle, this time changes little between $M=2$ and $M=6$. For $M=1$, the SQP starts from an initial guess instead of a previous solution, and the line search needs more trial points. The Hessian computation adds \SI{0.11}{ms} per iteration on average. Fig.~\ref{fig:cost_lower} shows that a lower-level call that returns $V_i$ and $\nabla_{t_i}V_i$ takes \SI{0.79}{ms}, and that the Hessian adds \SI{0.56}{ms} to the same solve. This addition consists of the factorization of the KKT matrix, the derivative of the KKT conditions with respect to $t_i$, which forms the right-hand side of the sensitivity system, and two substitutions with the factorization, one per crossing time, together with the transfer of the results to Python. Since each stage has only four states and one input, a substitution costs about one quarter of the factorization. The prediction QP, which only reuses derivatives already computed, is therefore much cheaper than a confirmation.

\subsection{Dense versus structured upper-level QP}
\label{sec:res_qp}

\begin{figure}[!t]
\centering
\includegraphics[width=\columnwidth]{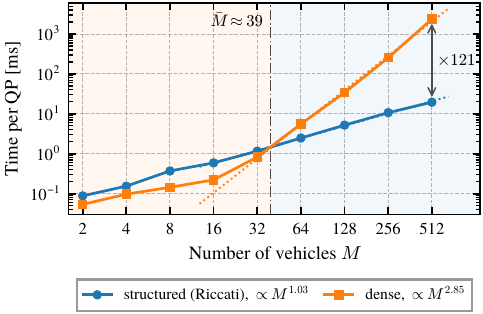}
\caption{Time per upper-level QP for the dense and the structured \texttt{HPIPM} formulations versus the number of vehicles $M$. Dotted lines: fitted power laws.}
\label{fig:scaling}
\end{figure}

Fig.~\ref{fig:scaling} compares the two \texttt{HPIPM} formulations of Section~\ref{sec:sqp} on upper-level QPs with $M=2$ to $512$ vehicles. The structured formulation scales as $M^{1.03}$, in line with the linear cost of the Riccati recursion, while the dense one scales as $M^{2.85}$. The dense formulation is faster for few vehicles. In each interior-point iteration, it factorizes a single matrix of size $2M$, which is cheap when $M$ is small. The structured formulation instead runs a Riccati recursion over $M$ stages, and each stage requires its own operations on matrices of size at most $2$. For few vehicles, this fixed work per stage costs more than the single dense factorization, whose cost grows as $M^3$. The two curves cross at $\bar M\approx39$, and at $M=512$ the structured formulation is $121$ times faster. The implementation therefore uses the dense formulation for $M<\bar M$ and the structured one otherwise.

\subsection{Closed-loop operation under computation delay}
\label{sec:res_closed}

The MPC scheme of \cite{shahi2026mpc} neglects the computational time $t_{\mathrm{alg}}$ needed for the schedule refinement to terminate. Because that assumption is unrealistic, any practical algorithm needs to account for $t_{\mathrm{alg}}$ and, once triggered, predict the initial states $\hat x_{i,0}$ after time $t_{\mathrm{alg}}$ and compute the schedule using these predictions rather than the current state measurements, which will be outdated once the algorithm terminates. Because different algorithms have different computation times $t_{\mathrm{alg}}$, we evaluate next which of the controllers proposed above performs best, i.e., is it worth adopting an approximate solution earlier or adopting the fully converged solution after a longer time?

Each run simulates \SI{300}{s} of traffic with more than 300 vehicles in five random seeds, light, medium and heavy in the ratio $0.4/0.2/0.4$, so that the intersection operates close to its capacity. All controllers store every evaluated pair in the memory $\mathcal{T}$, as in H1 of Section~\ref{sec:heuristic}.

Compared with the MPC scheme of \cite{shahi2026mpc}, we make the following changes. Each controller has a fixed delay of $\kappa=\lceil t_{\mathrm{alg}}^{\max}/t_\mathrm{s}\rceil$ sampling periods, where $t_{\mathrm{alg}}^{\max}$ is its largest computation time in calibration runs with the same traffic. A decision started at instant $k$ uses the states predicted for instant $k+\kappa$ and takes effect at $k+\kappa$, when the next decision starts. Until then, the previous schedule remains in force. A decision that takes longer than $\kappa\,t_\mathrm{s}$ is discarded. The crossing times of a vehicle are fixed once it can no longer stop before the intersection.
The four controllers share the lower and upper levels and differ only in the schedule refinement. C0 runs Algorithm~\ref{alg:swap} at every sampling instant with $\kappa=0$. It cannot be implemented and serves as the reference. C1 performs the same search but stops after \SI{2}{s}. C2 replans and evaluates at most one candidate per decision, namely the top-ranked pair not in $\mathcal{T}$, by a re-solve. C3 is the real-time variant of Section~\ref{sec:realtime}. The calibration gives $\kappa=28$, $13$ and $1$ for C1, C2 and C3.

\begin{figure}[!t]
\centering
\includegraphics[width=\columnwidth]{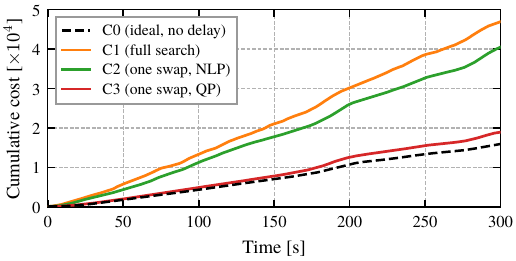}
\caption{Mean cumulative closed-loop cost over five runs of \SI{300}{s}.}
\label{fig:closed_loop}
\end{figure}

\begin{figure}[!t]
\centering
\includegraphics[width=\columnwidth]{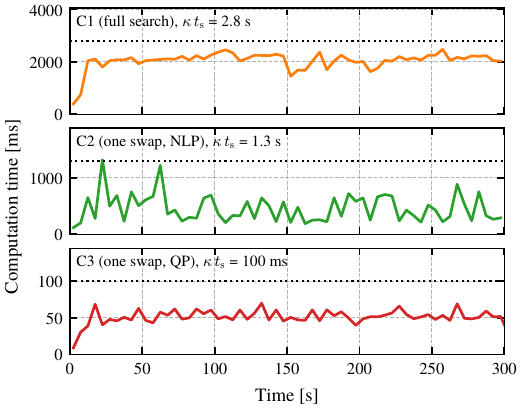}
\caption{Worst computation time per decision in each \SI{5}{s} window over five runs. Dotted: fixed delay $\kappa\,t_\mathrm{s}$.}
\label{fig:compute}
\end{figure}

Table~\ref{tab:closedloop} and Figs.~\ref{fig:closed_loop} and~\ref{fig:compute} summarize the results. In every run, C3 has the lowest cost after C0, followed by C2 and C1. C0 evaluates $3.9$ candidate swaps per decision on average and C1 $16.2$. C1 and C2 compute their decisions for states \SI{2.8}{s} and \SI{1.3}{s} ahead, and their costs are $197.2\%$ and $156.3\%$ above that of C0. The computation time of C2 varies strongly, from \SI{0.16}{s} on average to \SI{1.3}{s} at most, and since the fixed delay must cover the worst case, most of its decisions are computed for states \SI{1.3}{s} ahead. C3 decides at every sampling instant with a computation time of at most \SI{69.2}{ms}, and its cost is $19.7\%$ above that of C0. Its mean computation time is \SI{25.6}{ms}, of which \SI{17.1}{ms} are spent on the derivatives. In our simulation, these are evaluated for one vehicle after the other on a single thread, at \SI{0.75}{ms} per vehicle. In the architecture of Fig.~\ref{fig:architecture}, each vehicle evaluates only its own derivatives, so that a decision of C3 would take about \SI{9}{ms}: the two QPs and the acceptance step of the coordinator, which take \SI{8.2}{ms}, plus the derivatives of one vehicle.
\begin{table}[!t]
\centering
\caption{Closed-loop results, mean over five runs of \SI{300}{s}}
\label{tab:closedloop}
\setlength{\tabcolsep}{4pt}
\begin{tabular}{lrrrr}
\toprule
 & Cost increase & Fixed delay & \multicolumn{2}{c}{Comp.\ time [ms]} \\
\cmidrule(lr){4-5}
Controller & w.r.t.\ C0 [\%] & $\kappa\,t_\mathrm{s}$ [ms] & mean & max \\
\midrule
C0 (ideal)          & 0.0   & 0      & 449.5    & 10\,321.3 \\
C1 (full search)    & 197.2 & 2\,800 & 1\,483.6 & 2\,454.1 \\
C2 (one swap, NLP)  & 156.3 & 1\,300 & 162.8    & 1\,313.1 \\
C3 (one swap, QP)   & 19.7  & 100    & 25.6     & 69.2 \\
\bottomrule
\end{tabular}
\end{table}

\section{Conclusion}
\label{sec:conclusion}

We have presented a sensitivity-based SQP method for the coordination of CAVs at an intersection, built on the bi-level formulation of the coordination problem. The gradient and the Hessian of each vehicle's value function are obtained analytically from the \texttt{acados} solution sensitivities. With these derivatives, the bi-level fromulation yields the same crossing orders as the centralized formulation in $100$ randomized scenarios, and a single prediction QP predicts the cost change of a candidate swap with the correct sign in every confirmed case. The structured \texttt{HPIPM} formulation makes the upper-level QP scale linearly with the number of vehicles. In closed loop with a fixed computation delay, the real-time variant is the only controller that decides within the \SI{100}{ms} sampling period, and its cost is $19.7\%$ above that of the ideal controller without delay, whereas the controllers based on full re-solves lose $156.3\%$ and $197.2\%$ because of their latency. This work completes the line of research of \cite{shahi2026lagrange,shahi2026mpc}: it retains the multiplier guided swapping heuristic, replaces the expensive re-solves with sensitivity-based QPs, and shows that, once the computation time is accounted for, a fast approximate decision is worth more than an exact but late one.

\bibliographystyle{IEEEtran}
\bibliography{references}

@IEEEtranBSTCTL{IEEEexample:BSTcontrol,
  CTLdash_repeated_names = "no"
}

@article{riostorres_survey,
  author  = {Rios-Torres, Jackeline and Malikopoulos, Andreas A.},
  title   = {A survey on the coordination of connected and automated vehicles at intersections and merging at highway on-ramps},
  journal = {IEEE Transactions on Intelligent Transportation Systems},
  year    = {2017},
  volume  = {18},
  number  = {5},
  pages   = {1066--1077},
  doi     = {10.1109/TITS.2016.2600504}
}

@article{campos_mag,
  author  = {Rodrigues de Campos, Gabriel and Falcone, Paolo and Hult, Robert and Wymeersch, Henk and Sj{\"o}berg, Jonas},
  title   = {Traffic coordination at road intersections: Autonomous decision-making algorithms using model-based heuristics},
  journal = {IEEE Intelligent Transportation Systems Magazine},
  year    = {2017},
  volume  = {9},
  number  = {1},
  pages   = {8--21},
  doi     = {10.1109/MITS.2016.2630585}
}

@article{hult_theory,
  author  = {Hult, Robert and Zanon, Mario and Gros, S{\'e}bastien and Falcone, Paolo},
  title   = {Optimal coordination of automated vehicles at intersections: Theory and experiments},
  journal = {IEEE Transactions on Control Systems Technology},
  year    = {2019},
  volume  = {27},
  number  = {6},
  pages   = {2510--2525},
  doi     = {10.1109/TCST.2018.2871397}
}

@article{fayazi,
  author  = {Fayazi, Seyed Alireza and Vahidi, Ardalan},
  title   = {Mixed-integer linear programming for optimal scheduling of autonomous vehicle intersection crossing},
  journal = {IEEE Transactions on Intelligent Vehicles},
  year    = {2018},
  volume  = {3},
  number  = {3},
  pages   = {287--299},
  doi     = {10.1109/TIV.2018.2843163}
}

@article{colombo,
  author  = {Colombo, Alessandro and Del Vecchio, Domitilla},
  title   = {Least restrictive supervisors for intersection collision avoidance: A scheduling approach},
  journal = {IEEE Transactions on Automatic Control},
  year    = {2015},
  volume  = {60},
  number  = {6},
  pages   = {1515--1527},
  doi     = {10.1109/TAC.2014.2381453}
}

@article{dresner_stone,
  author  = {Dresner, Kurt and Stone, Peter},
  title   = {A multiagent approach to autonomous intersection management},
  journal = {Journal of Artificial Intelligence Research},
  year    = {2008},
  volume  = {31},
  pages   = {591--656},
  doi     = {10.1613/jair.2502}
}

@inproceedings{qian_priority,
  author    = {Qian, Xiangjun and Gregoire, Jean and Moutarde, Fabien and De La Fortelle, Arnaud},
  title     = {Priority-based coordination of autonomous and legacy vehicles at intersection},
  booktitle = {Proc. 17th International IEEE Conference on Intelligent Transportation Systems (ITSC)},
  year      = {2014},
  pages     = {1166--1171},
  doi       = {10.1109/ITSC.2014.6957845}
}

@misc{shahi2026lagrange,
  author       = {Shahi, Saugat and Fabiani, Filippo and Zanon, Mario},
  title        = {A Heuristic for Scheduling Autonomous Vehicles at
Intersections},
  year         = {2026},
  howpublished = {arXiv:2610.37275 [eess.SY]},
  url          = {https://arxiv.org/abs/2610.37275}
}

@inproceedings{hult_primal,
  author    = {Hult, Robert and Zanon, Mario and Gros, S{\'e}bastien and Falcone, Paolo},
  title     = {Primal decomposition of the optimal coordination of vehicles at traffic intersections},
  booktitle = {Proc. 55th IEEE Conference on Decision and Control (CDC)},
  year      = {2016},
  pages     = {2567--2573},
  doi       = {10.1109/CDC.2016.7798649}
}

@article{hult2020,
  author  = {Hult, Robert and Zanon, Mario and Frison, Gianluca and Gros, S{\'e}bastien and Falcone, Paolo},
  title   = {Experimental validation of a semi-distributed sequential quadratic programming method for optimal coordination of automated vehicles at intersections},
  journal = {Optimal Control Applications and Methods},
  year    = {2020},
  volume  = {41},
  number  = {4},
  pages   = {1068--1096}
}

@article{hult_ip,
  author  = {Hult, Robert and Zanon, Mario and Gros, S{\'e}bastien and Falcone, Paolo},
  title   = {A semidistributed interior point algorithm for optimal coordination of automated vehicles at intersections},
  journal = {IEEE Transactions on Control Systems Technology},
  year    = {2022},
  volume  = {30},
  number  = {5},
  pages   = {1977--1989},
  doi     = {10.1109/TCST.2021.3132835}
}

@article{malikopoulos_auto,
  author  = {Malikopoulos, Andreas A. and Cassandras, Christos G. and Zhang, Yue J.},
  title   = {A decentralized energy-optimal control framework for connected automated vehicles at signal-free intersections},
  journal = {Automatica},
  volume  = {93},
  pages   = {244--256},
  year    = {2018},
  doi     = {10.1016/j.automatica.2018.03.056}
}

@article{zhang_cassandras,
  author  = {Zhang, Yue and Cassandras, Christos G.},
  title   = {Decentralized optimal control of connected automated vehicles at signal-free intersections including comfort-constrained turns and safety guarantees},
  journal = {Automatica},
  volume  = {109},
  pages   = {108563},
  year    = {2019},
  doi     = {10.1016/j.automatica.2019.108563}
}

@article{malikopoulos_time,
  author  = {Malikopoulos, Andreas A. and Beaver, Logan and Chremos, Ioannis Vasileios},
  title   = {Optimal time trajectory and coordination for connected and automated vehicles},
  journal = {Automatica},
  volume  = {125},
  pages   = {109469},
  year    = {2021},
  doi     = {10.1016/j.automatica.2020.109469}
}

@article{kim_kumar,
  author  = {Kim, Kyoung-Dae and Kumar, Panganamala Ramana},
  title   = {An {MPC}-based approach to provable system-wide safety and liveness of autonomous ground traffic},
  journal = {IEEE Transactions on Automatic Control},
  year    = {2014},
  volume  = {59},
  number  = {12},
  pages   = {3341--3356}
}

@inproceedings{campos_rhc,
  author    = {Campos, Gabriel R. and Falcone, Paolo and Wymeersch, Henk and Hult, Robert and Sj{\"o}berg, Jonas},
  title     = {Cooperative receding horizon conflict resolution at traffic intersections},
  booktitle = {Proc. 53rd IEEE Conference on Decision and Control (CDC)},
  year      = {2014},
  pages     = {2932--2937},
  doi       = {10.1109/CDC.2014.7039840}
}

@article{katriniok,
  author  = {Katriniok, Alexander and Kleibaum, Peter and Jo{\v{s}}evski, Martina},
  title   = {Distributed model predictive control for intersection automation using a parallelized optimization approach},
  journal = {IFAC-PapersOnLine},
  volume  = {50},
  number  = {1},
  pages   = {5940--5946},
  year    = {2017},
  doi     = {10.1016/j.ifacol.2017.08.1492}
}

@article{xu_distributed,
  author  = {Xu, Biao and Li, Shengbo Eben and Bian, Yougang and Li, Shen and Ban, Xuegang Jeff and Wang, Jianqiang and Li, Keqiang},
  title   = {Distributed conflict-free cooperation for multiple connected vehicles at unsignalized intersections},
  journal = {Transportation Research Part C: Emerging Technologies},
  volume  = {93},
  pages   = {322--334},
  year    = {2018},
  doi     = {10.1016/j.trc.2018.06.004}
}

@inproceedings{murgovski,
  author    = {Murgovski, Nikolce and de Campos, Gabriel Rodrigues and Sj{\"o}berg, Jonas},
  title     = {Convex modeling of conflict resolution at traffic intersections},
  booktitle = {Proc. 54th IEEE Conference on Decision and Control (CDC)},
  year      = {2015},
  pages     = {4708--4713},
  doi       = {10.1109/CDC.2015.7402953}
}

@inproceedings{qian_mpc,
  author    = {Qian, Xiangjun and Gregoire, Jean and de La Fortelle, Arnaud and Moutarde, Fabien},
  title     = {Decentralized model predictive control for smooth coordination of automated vehicles at intersection},
  booktitle = {Proc. European Control Conference (ECC)},
  year      = {2015},
  pages     = {3452--3458},
  doi       = {10.1109/ECC.2015.7331068}
}

@inproceedings{kneissl,
  author    = {Kneissl, Maximilian and Molin, Adam and Esen, Hasan and Hirche, Sandra},
  title     = {A feasible {MPC}-based negotiation algorithm for automated intersection crossing},
  booktitle = {Proc. European Control Conference (ECC)},
  year      = {2018},
  pages     = {1282--1288},
  doi       = {10.23919/ECC.2018.8550251}
}

@article{chalaki,
  author  = {Chalaki, Behdad and Malikopoulos, Andreas A.},
  title   = {Optimal control of connected and automated vehicles at multiple adjacent intersections},
  journal = {IEEE Transactions on Control Systems Technology},
  year    = {2022},
  volume  = {30},
  number  = {3},
  pages   = {972--984},
  doi     = {10.1109/TCST.2021.3082306}
}

@article{ge_graphical,
  author  = {Ge, Qiang and Sun, Qi and Wang, Zhen and Li, Shengbo Eben and Gu, Ziqing and Zheng, Sifa and Liao, Lyuchao},
  title   = {Real-time coordination of connected vehicles at intersections using graphical mixed integer optimization},
  journal = {IET Intelligent Transport Systems},
  volume  = {15},
  number  = {6},
  pages   = {795--807},
  year    = {2021},
  doi     = {10.1049/itr2.12061}
}

@inproceedings{hult_miqp,
  author    = {Hult, Robert and Zanon, Mario and Gros, S{\'e}bastien and Falcone, Paolo},
  title     = {An {MIQP}-based heuristic for optimal coordination of vehicles at intersections},
  booktitle = {Proc. 57th IEEE Conference on Decision and Control (CDC)},
  year      = {2018},
  pages     = {2783--2790},
  doi       = {10.1109/CDC.2018.8618945}
}

@article{xu_tree,
  author  = {Xu, Huile and Zhang, Yi and Li, Li and Li, Weixia},
  title   = {Cooperative driving at unsignalized intersections using tree search},
  journal = {IEEE Transactions on Intelligent Transportation Systems},
  year    = {2020},
  volume  = {21},
  number  = {11},
  pages   = {4563--4571},
  doi     = {10.1109/TITS.2019.2940641}
}

@article{li_zhou,
  author  = {Li, Peng-Fei and Zhou, Xuesong},
  title   = {Recasting and optimizing intersection automation as a connected-and-automated-vehicle ({CAV}) scheduling problem: A sequential branch-and-bound search approach in phase-time-traffic hypernetwork},
  journal = {Transportation Research Part B: Methodological},
  year    = {2017},
  volume  = {105},
  pages   = {479--506}
}

@inproceedings{yan,
  author    = {Yan, Fei and Dridi, Mahjoub and El Moudni, Abdellah},
  title     = {Autonomous vehicle sequencing algorithm at isolated intersections},
  booktitle = {Proc. 12th International IEEE Conference on Intelligent Transportation Systems (ITSC)},
  year      = {2009},
  pages     = {1--6},
  doi       = {10.1109/ITSC.2009.5309708}
}

@article{faris_reorder,
  author  = {Faris, Muhammad and Zanon, Mario and Falcone, Paolo},
  title   = {An optimization-based dynamic reordering heuristic for coordination of vehicles in mixed traffic intersections},
  journal = {IEEE Transactions on Control Systems Technology},
  year    = {2025},
  volume  = {33},
  number  = {4},
  pages   = {1387--1402},
  doi     = {10.1109/TCST.2024.3508542}
}

@misc{shahi2026mpc,
  author       = {Shahi, Saugat and Fabiani, Filippo and Zanon, Mario},
  title        = {Dynamic Coordination of Autonomous Vehicles at Intersections},
  year         = {2026},
  howpublished = {arXiv:2610.37288 [eess.SY]},
  url          = {https://arxiv.org/abs/2610.37288}
}

@book{rawlings_mpc,
  author    = {Rawlings, James B. and Mayne, David Q. and Diehl, Moritz M.},
  title     = {Model Predictive Control: Theory, Computation, and Design},
  edition   = {2nd},
  publisher = {Nob Hill Publishing},
  address   = {Madison, WI, USA},
  year      = {2017}
}

@article{ipopt,
  author  = {W{\"a}chter, Andreas and Biegler, Lorenz T.},
  title   = {On the implementation of an interior-point filter line-search algorithm for large-scale nonlinear programming},
  journal = {Mathematical Programming},
  year    = {2006},
  volume  = {106},
  number  = {1},
  pages   = {25--57},
  doi     = {10.1007/s10107-004-0559-y}
}

@article{casadi,
  author  = {Andersson, Joel A. E. and Gillis, Joris and Horn, Greg and Rawlings, James B. and Diehl, Moritz},
  title   = {{CasADi}: a software framework for nonlinear optimization and optimal control},
  journal = {Mathematical Programming Computation},
  year    = {2019},
  volume  = {11},
  number  = {1},
  pages   = {1--36},
  doi     = {10.1007/s12532-018-0139-4}
}

@book{fiacco,
  author    = {Fiacco, Anthony V.},
  title     = {Introduction to Sensitivity and Stability Analysis in Nonlinear Programming},
  year      = {1983},
  publisher = {Academic Press},
  address   = {New York, NY, USA},
  series    = {Mathematics in Science and Engineering},
  volume    = {165}
}

@incollection{buskens_maurer,
  author    = {B{\"u}skens, Christof and Maurer, Helmut},
  title     = {Sensitivity analysis and real-time optimization of parametric nonlinear programming problems},
  booktitle = {Online Optimization of Large Scale Systems},
  editor    = {Gr{\"o}tschel, Martin and Krumke, Sven O. and Rambau, J{\"o}rg},
  publisher = {Springer},
  address   = {Berlin, Heidelberg},
  year      = {2001},
  pages     = {3--16},
  doi       = {10.1007/978-3-662-04331-8_1}
}

@article{zavala_biegler,
  author  = {Zavala, Victor M. and Biegler, Lorenz T.},
  title   = {The advanced-step {NMPC} controller: Optimality, stability and robustness},
  journal = {Automatica},
  volume  = {45},
  number  = {1},
  pages   = {86--93},
  year    = {2009},
  doi     = {10.1016/j.automatica.2008.06.011}
}

@article{rti,
  author  = {Diehl, Moritz and Bock, Hans Georg and Schl{\"o}der, Johannes P.},
  title   = {A real-time iteration scheme for nonlinear optimization in optimal feedback control},
  journal = {SIAM Journal on Control and Optimization},
  year    = {2005},
  volume  = {43},
  number  = {5},
  pages   = {1714--1736},
  doi     = {10.1137/S0363012902400713}
}

@article{acados,
  author  = {Verschueren, Robin and Frison, Gianluca and Kouzoupis, Dimitris and Frey, Jonathan and van Duijkeren, Niels and Zanelli, Andrea and Novoselnik, Branimir and Albin, Thivaharan and Quirynen, Rien and Diehl, Moritz},
  title   = {acados---a modular open-source framework for fast embedded optimal control},
  journal = {Mathematical Programming Computation},
  year    = {2022},
  volume  = {14},
  pages   = {147--183},
  doi     = {10.1007/s12532-021-00208-8}
}

@misc{frey2025differentiable,
  author       = {Frey, Jonathan and Baumg{\"a}rtner, Katrin and Frison, Gianluca and Reinhardt, Dirk and Hoffmann, Jasper and Fichtner, Leonard and Gros, Sebastien and Diehl, Moritz},
  title        = {Differentiable nonlinear model predictive control},
  howpublished = {arXiv preprint arXiv:2505.01353},
  year         = {2025}
}

@article{hpipm,
  author  = {Frison, Gianluca and Diehl, Moritz},
  title   = {{HPIPM}: a high-performance quadratic programming framework for model predictive control},
  journal = {IFAC-PapersOnLine},
  volume  = {53},
  number  = {2},
  pages   = {6563--6569},
  year    = {2020},
  doi     = {10.1016/j.ifacol.2020.12.073}
}

@book{nocedal,
  author    = {Nocedal, Jorge and Wright, Stephen J.},
  title     = {Numerical Optimization},
  edition   = {2nd},
  year      = {2006},
  publisher = {Springer},
  address   = {New York, NY, USA},
  doi       = {10.1007/978-0-387-40065-5}
}

@inproceedings{frison_condensing,
  author    = {Frison, Gianluca and Kouzoupis, Dimitris and J{\o}rgensen, John Bagterp and Diehl, Moritz},
  title     = {An efficient implementation of partial condensing for nonlinear model predictive control},
  booktitle = {Proc. 55th IEEE Conference on Decision and Control (CDC)},
  pages     = {4457--4462},
  year      = {2016},
  doi       = {10.1109/CDC.2016.7798946}
}

\end{document}